\documentclass[ioprevtex4]{emulateapj}
\usepackage{graphicx}				
\usepackage{amssymb}
\usepackage{color}
\usepackage{amsmath}
\usepackage{tablefootnote}

\usepackage{natbib}
\shortauthors{Forrest et al.}
\shorttitle{The IRX$-\beta$ Relation From $\sim$4000 Galaxies}

\begin{document}
\title{UV to IR Luminosities and Dust Attenuation Determined from $\sim$4000 K-Selected Galaxies at $1<\MakeLowercase{z}<3$ in the ZFOURGE Survey\altaffilmark{*}}
\author{
Ben Forrest\altaffilmark{1,2},  		
Kim-Vy H. Tran\altaffilmark{2},  		
Adam R. Tomczak\altaffilmark{2,3},  	
Adam Broussard\altaffilmark{2}, 	
Ivo Labb\'e\altaffilmark{4},	 		
Casey Papovich\altaffilmark{2}, 	
Mariska Kriek\altaffilmark{5},		
Rebecca J. Allen\altaffilmark{6},	
Michael Cowley\altaffilmark{7,8},	
Mark Dickinson\altaffilmark{9},		
Karl Glazebrook\altaffilmark{6},		
Josha van Houdt\altaffilmark{4},	
Hanae Inami\altaffilmark{9},		
Glenn G. Kacprzak\altaffilmark{6},	
Lalitwadee Kawinwanichakij\altaffilmark{2},	
Daniel Kelson\altaffilmark{10},		
Patrick J. McCarthy\altaffilmark{10},	
Andrew Monson\altaffilmark{10},	
Glenn Morrison\altaffilmark{11,12},	
Themiya Nanayakkara\altaffilmark{6},	
S. Eric Persson\altaffilmark{10},	
Ryan F. Quadri\altaffilmark{2}, 		
Lee R. Spitler\altaffilmark{7,8},		
Caroline Straatman\altaffilmark{4},	
Vithal Tilvi\altaffilmark{2,13}		
}

\altaffiltext{*}{This paper includes data gathered with the 6.5 meter Magellan Telescopes located at Las Campanas Observatory, Chile.}
\altaffiltext{1}{bforrest@physics.tamu.edu}
\altaffiltext{2}{George P. and Cynthia W. Mitchell Institute for Fundamental Physics and Astronomy, Department of Physics and Astronomy, Texas A\&M University, College Station, TX 77843, USA}
\altaffiltext{3}{Department of Physics, UC Davis, Davis, CA 95616, USA}
\altaffiltext{4}{Leiden Observatory, Leiden University, PO Box 9513, 2300 RA Leiden, The Netherlands}
\altaffiltext{5}{Astronomy Department, University of California at Berkeley, Berkeley, CA 94720, USA}
\altaffiltext{6}{Centre for Astrophysics and Supercomputing, Swinburne University, Hawthorn, VIC 3122, Australia}
\altaffiltext{7}{Australian Astronomical Observatory, PO Box 915, North Ryde, NSW 1670, Australia}
\altaffiltext{8}{Department of Physics \& Astronomy, Macquarie University, Sydney, NSW 2109, Australia}
\altaffiltext{9}{National Optical Astronomy Observatory, 950 North Cherry Avenue, Tucson, AZ 85719, USA}
\altaffiltext{10}{Carnegie Observatories, Pasadena, CA 91101, USA}
\altaffiltext{11}{Institute for Astronomy, University of Hawaii, Honolulu, Hawaii, HI-96743, USA}
\altaffiltext{12}{Canada-France-Hawaii Telescope, Kamuela, Hawaii, HI-96743, USA}
\altaffiltext{13}{School of Earth \& Space Exploration, Arizona State University, Tempe, AZ 85287}

\begin{abstract}
We build a set of composite galaxy SEDs by de-redshifting and scaling multi-wavelength photometry from galaxies in the ZFOURGE survey, covering the CDFS, COSMOS, and UDS fields. 
From a sample of $\sim$4000 $K_s$-band selected galaxies, we define 38 composite galaxy SEDs that yield continuous low-resolution spectra $(R\sim 45)$ over the rest-frame range 0.1-4 $\mu m$.
Additionally, we include far infrared photometry from the \textit{Spitzer Space Telescope} and the \textit{Herschel Space Observatory} to characterize the infrared properties of our diverse set of composite SEDs.
From these composite SEDs we analyze the rest-frame \textit{UVJ} colors, as well as the ratio of IR to UV light (IRX) and the UV slope ($\beta$) in the IRX$-\beta$ dust relation at $1<z<3$.
Blue star-forming composite SEDs show IRX and $\beta$ values consistent with local relations; dusty star-forming galaxies have considerable scatter, as found for local IR bright sources, but on average appear bluer than expected for their IR fluxes.
We measure a tight linear relation between rest-frame \textit{UVJ} colors and dust attenuation for star-forming composites, providing a direct method for estimating dust content from either $(U-V)$ or $(V-J)$ rest-frame colors for star-forming galaxies at intermediate redshifts.
\end{abstract}


\section{Introduction} \label{Intro}

Constraining the dust content of galaxies is vital to improving our knowledge of star formation histories and galaxy evolution.
The geometry and orientation of dust grains, as well as their spatial distribution in galaxies greatly affect the shape of a galaxy's observed spectral energy distribution (SED) \citep[e.g.,][]{Chevallard2013, Casey2014, Penner2015, Salmon2015}.
Correcting for these effects is necessary to understand the intrinsic properties of a galaxy.
For nearby galaxies, spectroscopy provides insight into these effects.
However, because spectroscopy requires significantly more telescope time and brighter targets than photometry, photometry is a better choice for large, deep samples.

Beyond the local universe, galactic properties are often determined by fitting stellar population synthesis models to a handful of photometric points, then assigning redshifts, ages, metallicities, masses, etc. to the galaxy based on the best fitting SED \citep[e.g.,][]{Papovich2001,Franx2003}.
The use of medium band near-IR filters in the NEWFIRM Medium Band Survey \citep[NMBS;][]{Whitaker2011} and the Fourstar Evolution Survey \citep[ZFOURGE;][submitted]{Straatman2016} has enabled more accurate photometric redshift ($z_{phot}$) measurements of large numbers of galaxies, on the order of $\sigma_{NMAD}\sim0.01-0.02$ (for $K_s<25$) when compared to higher quality redshifts from grism \citep{Bezanson2015} and spectroscopic observations \citep{Nanayakkara2016}.
These surveys also cover legacy fields which have extensive photometric observations from the rest-frame UV to the near-IR.

The combination of greater photometric sampling and near-IR filters which identify the 4000 $\textrm{\AA}$ break allow for a more constrained SED fit \citep[e.g.,][]{Kriek2013a}.
This makes these surveys prime datasets for the development of multi-wavelength composite SEDs.
If one can determine which galaxies have intrinsically similar SEDs, then by de-redshifting and scaling photometry for galaxies at several redshifts one can generate a well-sampled composite SED.
Over the last few years, several papers have demonstrated the effectiveness of multi-wavelength composite SEDs in determining galaxy properties \citep[e.g.,][]{Kriek2011, Kriek2013a, Utomo2014, Yano2016}.
In this work we use data from the ZFOURGE survey (http://zfourge.tamu.edu/) to define composite SEDs due not only to the survey's accurate redshifts, but also its depth (limiting magnitude of $K\sim~25.5$ mag).
This allows us to build composite SEDs from galaxies at higher redshifts and lower masses than previous studies while still maintaining precision in our $z_{phot}$ measurements.

The optical to near-infrared (IR) SED characterizes the properties of the stellar populations. 
To better track total star formation, rest-frame mid-far IR observations, which indicate the amount of dust heated by young, massive stars, are essential \citep[e.g.,][]{Kennicutt1998, Kennicutt2012}.
The ultraviolet (UV) flux more directly traces these stars, and the ratio of these two components, the infrared excess (IRX), is a tracer of dust attenuation in the UV.
The UV slope ($\beta$) is also sensitive to the effects of dust \citep{Calzetti1994}, and it can be compared to the IRX to determine how dust attenuation affects the light of star-forming galaxies. 

This IRX$-\beta$ relation has been fit for various samples in the local universe \citep[e.g.,][]{Meurer1999a, Howell2010,Overzier2011} and in some cases compared to samples at higher redshifts \citep{Reddy2010, Reddy2011, Penner2012, Casey2014, Salmon2015}.
Notably, \cite{Howell2010} found that (U)LIRGs in the local universe do show significant scatter about IRX$-\beta$ relations, largely due to variations in IR flux.
The resulting relations have also been used to derive properties such as continuum reddening \citep{Puglisi2015} and distributions of dust in dust-obscured galaxies (DOGs) \citep{Penner2012}.
Several of these works found discrepancies between high-$z$ dusty star-forming galaxies and local IRX$-\beta$ relations.
Many of these studies estimate IR fluxes from a single photometric point, usually a \textit{Spitzer}/MIPS 24 $\mu m$ flux.
Our inclusion of \textit{Herschel} data broadens the IR wavelength range and improves determination of the IR flux, although uncertainties still do exist.

We assume a $\Lambda$CDM cosmology of $\Omega_M = 0.3$, $\Omega_\Lambda = 0.7$, and $H_0 = 70$ km s$^{-1}$ Mpc$^{-1}$ and a Chabrier IMF \citep{Chabrier2003}, and adopt an AB magnitude system \citep{Oke1983}.

\section{Data $\&$ Methods} \label{D&M}

\subsection{Data}
We use photometric data from the deep near-IR ZFOURGE survey \citep[][submitted]{Straatman2016}, covering the CDFS \citep{Giacconi2002}, COSMOS \citep{Scoville2007}, and UDS \citep{Lawrence2007} fields, as well as archival data to obtain photometric coverage over observed wavelengths ranging from 0.3 $\mu m$ to 8 $\mu m$.
The near-IR filters of ZFOURGE split the traditional $J$ and $H$ bands into 3 and 2 medium-band filters, respectively.
These allow us to constrain the photometric redshifts of the observed galaxies with a much higher precision than previously available - $\sigma_{NMAD} = 0.02$ \citep[][submitted]{Nanayakkara2016}.

We also include data from \textit{Spitzer}/MIPS 24 $\mu m$ (GOODS-S: PI Dickinson, COSMOS: PI Scoville, UDS: PI Dunlop) and \textit{Herschel}/PACS 100 $\mu m$ and 160 $\mu m$ filters from deep Herschel surveys of GOODS-S (Elbaz et al. 2011) and of the CANDELS COSMOS and UDS fields (PI: Dickinson; Inami et al. in preparation).
Section 2.3 from \cite{Tomczak2015} provides a description of how IR fluxes were measured.
Critically, these data allow better characterization of the rest-frame infrared wavelengths - and therefore dust content - of many of these galaxies.

To reduce any uncertainties due to photometric errors, we use a $K_s$-band signal to noise cut of 20.
We also restrict our sample to the redshift range $1.0<z<3.0$.


	\begin{figure*}[th]
	\centerline{\includegraphics[width=0.9\textwidth,trim=0in 1.2in 0in 1.1in, clip=true]{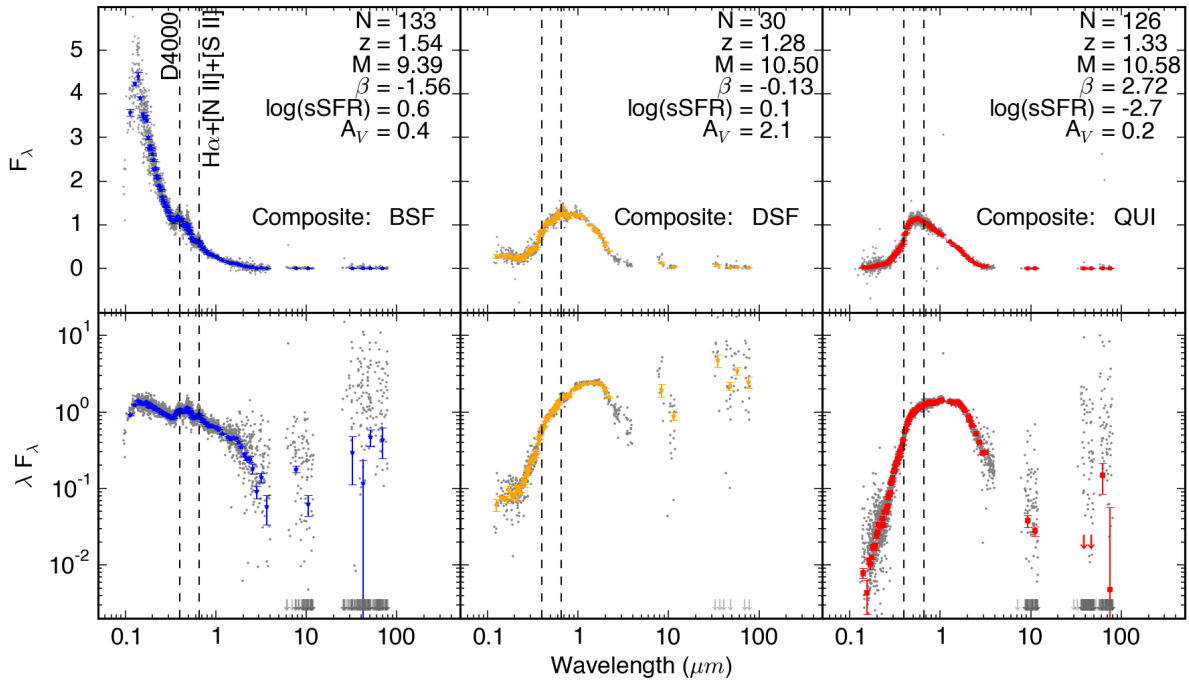}}
	\caption{Examples of our composite SEDs, including a blue star-forming composite SED (BSF), a dusty star-forming composite SED (DSF), and a quiescent composite SED (QUI).  Colored points represent the composite SEDs with NMAD scatter on the median as errorbars, while the gray points are the de-redshifted, scaled photometry from observations. Downward arrows show non-detections in the \textit{Spitzer}/MIPS and \textit{Herschel}/PACS filters and reflect the flux limits in those bands. The numbers given are the number of galaxies in the composite SED, the $z_{phot}$ and mass of the median galaxies, as well as the UV slope, the logarithm of the specific star formation rate (sSFR), and the dust attenuation $A_V$, also determined from SED-fitting.  Vertical dashed lines mark the location of the D4000 break and the H$\alpha$ line blend. The increased IR flux in the dusty star-forming composite SED relative to the quiescent composite SED likely reflects reddening from dust. This is also supported by the $A_V$ values reported by FAST. \vspace{20pt}}
	\label{fig: Composites}
	\end{figure*}


\subsection{Building Composite SEDs}
We outline our method, which broadly follows the methods of \cite{Kriek2011} and \cite{Utomo2014}, in this section.
More details will be included in an upcoming paper (Forrest et al., in prep).

After making the cuts mentioned above, we are left with 3984 galaxies.
Photometry in 22 artificial rest-frame filters is calculated for each galaxy as synthetic photometry based on an SED fit using EAZY \citep{Brammer2008}.
The similarity of galaxies is calculated using the shape of the synthetic photometry as in \cite{Kriek2011}:
\begin{align}
b_{12} &= \sqrt{\frac{\Sigma(f_{\lambda}^{ob1}-a_{12}f_{\lambda}^{ob2})^2}{\Sigma(f_{\lambda}^{ob1})^2}}\\
a_{12} &= \frac{\Sigma f_{\lambda}^{ob1}f_{\lambda}^{ob2}}{\Sigma (f_{\lambda}^{ob2}) ^2},
\end{align}
where $b_{12}$ is the (dis)similarity between SED shape as probed by the synthetic photometry, and $a_{12}$ is a scaling factor.
 
The galaxy with the most similar galaxies ($b<0.05$) is termed the primary, and those similar to it are termed analogs.
Once groupings are finalized, the observed photometry of all galaxies in a group is de-redshifted and scaled to unity in the optical.
Medians of these points are taken in bins of wavelength to construct a composite galaxy SED.
We obtain 38 composite SEDs from 2598 galaxies that we use for the remainder of this analysis; these have a resolving power of $R\sim 45$ in the rest-frame optical.
The remaining galaxies are in groups with fewer than 20 analogs; these galaxies will be explored in future work (Forrest et al., in prep).
Example composite SEDs are shown in Figure \ref{fig: Composites}.

Each composite point is dependent on the filter curves of the underlying photometry.
To determine rest-frame colors and parameters from SED-fitting, custom filter curves are defined for each point in our composite SEDs.
This is done by de-redshifting observed filter curves and scaling them to equal volume, then summing their responses.
Note that the errorbars on the composite SED points are $\sigma_{NMAD}$ errors on the medians and do not represent the errors in the photometry of the analogs.
In addition, a number of galaxies have negative flux measurements in the IR because their photometry is dominated by noise in the background-subtracted images; these are plotted as downward arrows in Figure \ref{fig: Composites}.
Such non-detections are included when calculating medians and errors to build our composites and are not removed.
These composite SEDs reveal details, such as H$\alpha$ emission, that are usually only available through spectroscopy.

For each of our composite SEDs, we also generate 100 bootstrapped composite SEDs.
Each one is made by performing a bootstrap resampling of the analog galaxies for the composite SED and recalculating the SED points and custom filter curves.
For parameters such as UV slope, UV flux, and IR flux, the same methods are applied to these bootstrapped composite SEDs to obtain errors on said parameters.

\section{Analysis} \label{Res}

\subsection{\textit{UVJ} Colors of Composite SEDs} \label{Res1}
\begin{figure}[tp]
	\centerline{\includegraphics[width=0.5\textwidth,trim=0in 0.2in 0in 0in, clip=true]{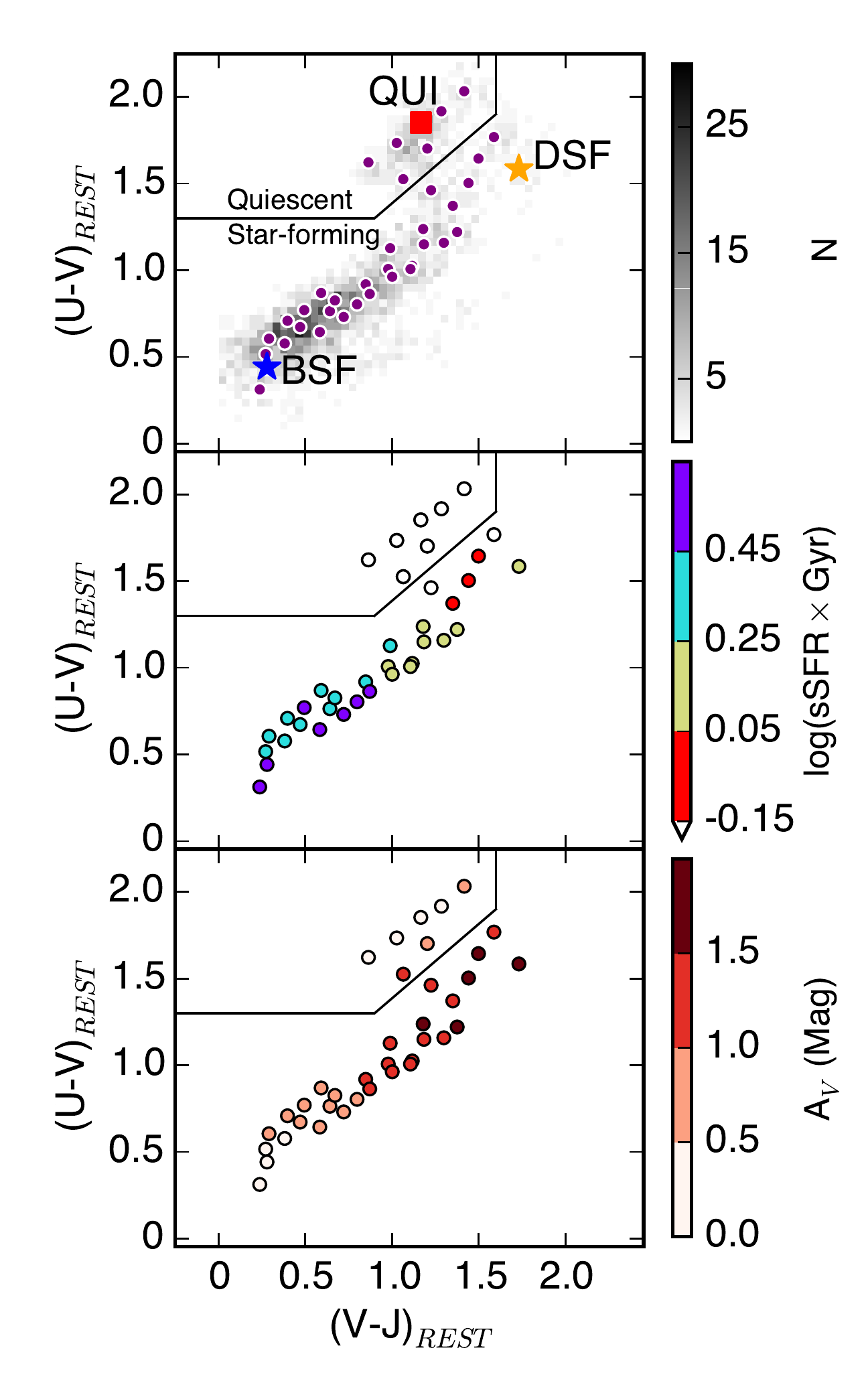}}
	\caption{Composite SEDs on the \textit{UVJ} diagram.  The top panel shows the distribution of the parent sample (greyscale) behind the composite SEDs (purple).  The example quiescent (red square), blue star-forming (blue star), and dusty star-forming (yellow star) composite SEDs shown in Figure \ref{fig: Composites} are labeled as well.  The middle and bottom panels show the composite SEDs colored by the logarithm of the sSFR and $A_V$ from SED-fitting, respectively. The composite SEDs show known trends with sSFR and $A_V$, as quiescent composite SEDs have much smaller sSFR values, and dusty star-forming composite SEDs have the greatest amount of dust attenuation.\vspace{20pt}}
	\label{fig: UVJ}
	\end{figure}

In the star-forming section of the \textit{UVJ} plane,  a strong correlation exists between redder colors and increased dust attenuation \citep{Wuyts2007,Williams2009,Patel2012}.
Additionally, the quiescent population has a considerably lower specific star formation rate (sSFR) \citep[][submitted]{Williams2009, Papovich2015, Straatman2016}.
It should be noted that the sSFR values reported by our SED-fitting program, FAST \citep{Kriek2009} do not consider the IR portion of the data. 
In Figure \ref{fig: UVJ} we plot our composite SEDs on the \textit{UVJ} diagram to analyze these relations.
The composite SEDs span the range covered by observed galaxies quite well, although the most extreme colors are not represented due to their rarity in our initial sample.
The previously known strong trend of star-forming galaxies with increasing dust ($A_V$) from the bottom-left to the top-right of the plot is quite clear, as is the strong decrease in sSFR from the star-forming to the quiescent regions.
We analyze these results in conjunction with our IRX$-\beta$ results in Section \ref{Res3}.

\subsection{IRX$-\beta$ Relation of Star-Forming Composite SEDs} \label{Res2}
To obtain $\beta$ we fit the scaled rest-frame UV data from our composite SEDs with a power law of the form $F_\lambda \propto \lambda^\beta$.
In this work we use photometry with rest-frame wavelengths in the range  $1500 < \lambda / \textrm{\AA} < 2600$ to fit the UV slope, similar to the range of the \cite{Calzetti1994} fitting windows, and mask points within 175 $\textrm{\AA}$ of the 2175 $\textrm{\AA}$ feature \citep{Noll2009, Buat2011, Buat2012}.
We also obtain the UV flux for our IRX calculation by integrating under the power law fit determined above in a 350 $\textrm{\AA}$ window centered on 1600 $\textrm{\AA}$ as in \cite{Meurer1999a}.

We calculate the IR flux by fitting the average template from \cite{Chary2001} to the scaled, de-redshifted composite SED IR points, and integrating under the resultant template from 8$-$1000 $\mu m$ (Forrest et al., in prep).
Additionally we only use such points that are at longer wavelengths than 8 $\mu m$ in the rest-frame.
Individual galaxies in our composite subsamples have luminosities $9.7 < \log (L_{IR}/L_\odot) < 13.6$, while the average of galaxies in a single composite range $10.5 < \log (L_{IR}/L_\odot) < 11.5$.
Galaxies in a particular composite have a median $L_{IR}$ scatter of 0.4 dex.
For star-forming composite SEDs, this model is sufficient for our analysis.
It should be noted however that the \cite{Chary2001} templates are not designed to fit quiescent galaxies.
As such, quiescent composite SEDs (as determined by position on the \textit{UVJ} diagram) are neither included on our IRX$-\beta$ plot, nor considered when calculating our IRX$-\beta$ relation.


	\begin{figure}[tp]
	\centerline{\includegraphics[width=0.5\textwidth,trim=0in 0.2in 0in 0in, clip=true]{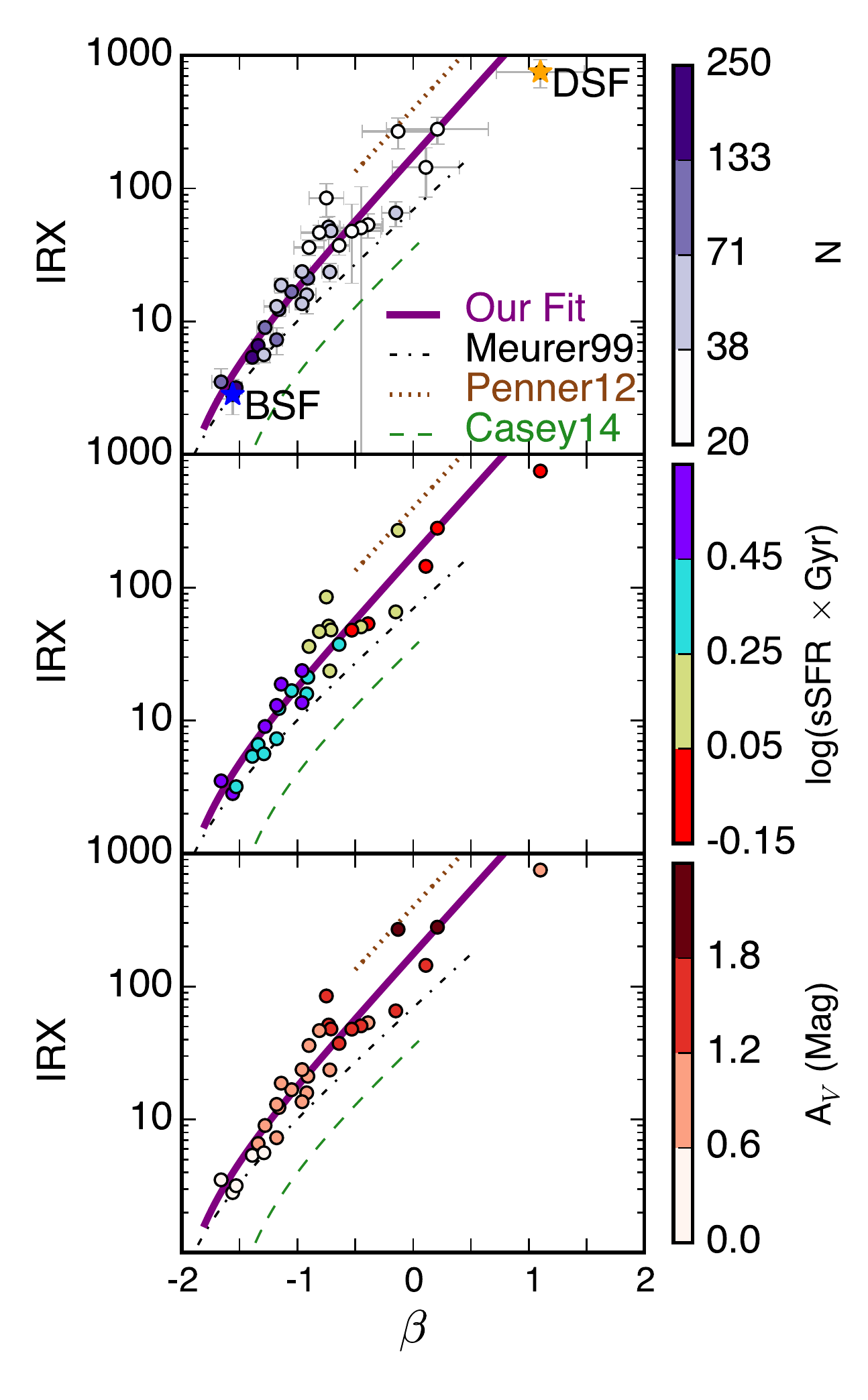}}
	\caption{Star-forming composite SEDs (as determined by position on the \textit{UVJ} diagram) on the IRX$-\beta$ diagram.  The point colors correspond to the number of analog galaxies (purple), the logarithm of the sSFR (rainbow), and the $A_V$ (red) from SED-fitting. 
	Also shown are two local fits - \cite{Meurer1999a} (black dashed-dotted line) and \cite{Casey2014} (green dashed line) - and two fits to $z\sim2$ data - this work (purple line) and a fit to the median points of \cite{Penner2012} (brown dotted line).
	Our dusty star-forming composite SEDs lie systematically above these local relations, appearing bluer than expected for their IR fluxes.\vspace{10pt}}
	\label{fig: L_IRXB}
	\end{figure}


Having obtained the three measurements necessary for the IRX$-\beta$ plot, we show our star-forming composite SEDs in Figure \ref{fig: L_IRXB}. 
We find the IRX$-\beta$ relation that fits these composite SEDs using the form
\begin{eqnarray}
IRX = BC_{UV} \times [10^{0.4A_{1600}}-1].
\end{eqnarray}
Here BC$_{UV}$ corrects to obtain all luminosity redward of the Lyman break (912 $\textrm{\AA}$).
We assume $BC_{UV} = 1.68$, as derived in similar studies \citep{Meurer1999a, Overzier2011, Takeuchi2012, Casey2014} to compute the least squares best fit to the data.

The other parameter is the dust attenuation at 1600 $\textrm{\AA}$, which is assumed to be a foreground screen, and thus linearly correlated with $\beta$ as $A_{1600} = q +r\beta$ \citep{Meurer1999a}. 
Performing a fit to the star-forming composite SEDs we obtain:
\begin{eqnarray}
IRX = 1.68 \times [10^{0.4(5.05+2.39\beta)}-1].
\end{eqnarray}


	\begin{figure*}[th]
	\centerline{\includegraphics[width=1\textwidth,trim=0in 0.2in 0.5in 0.2in]{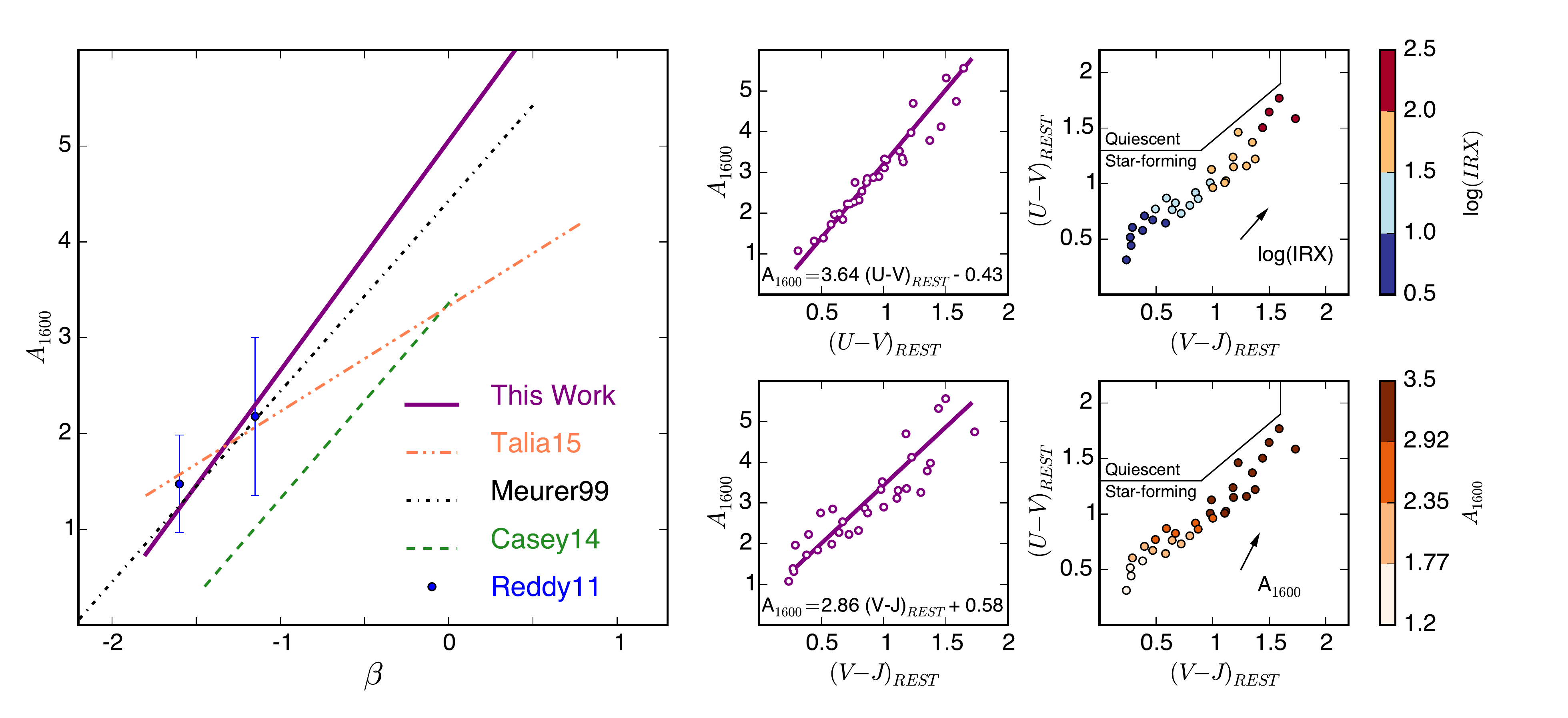}}
	\caption{A comparison of different extinction relations from literature $A_{1600}-\beta$ fits is shown on the left panel.  The $z\sim2$ LBG stacks from \cite{Reddy2011} are shown as blue points, which are consistent with several relations within the errorbars.  While galaxies with steeper UV slopes are in general agreement, our dusty star-forming composite SEDs show more extinction than would be determined from local relations.  Fit relations between our derived $A_{1600}$ and \textit{UVJ} rest-frame colors are shown in the middle column. These relations advocate the use of rest-frame optical colors to probe UV dust attenuation at these redshifts.  The right panels show the \textit{UVJ} diagrams color-coded by IRX and $A_{1600}$, with unit vectors for each parameter showing the effects of increasing the dust content. \vspace{20pt}}
	\label{fig: L_A1600}
	\end{figure*}


Our slope parameter of $r=2.39$ is steeper than previously determined values of $r\sim 2$ (see Table 1).
There is no discrepancy in our blue star-forming composite SEDs, as they are consistent with both local relations \citep[e.g.,][]{Meurer1999a,Overzier2011,Takeuchi2012} and higher redshift samples \citep[e.g.,][]{Reddy2010, Reddy2011}.
This can be seen in the left panel of Figure \ref{fig: L_A1600}.

Several previous studies of IR-luminous galaxies at $z\sim0$ have shown that IR-luminous galaxies have greater scatter in the IRX$-\beta$ plane and lie above relations derived for starburst galaxies \citep[e.g.][]{Howell2010, Overzier2011, Reddy2011,Casey2014}.
Similarly, the composite SEDs most discrepant from the \cite{Meurer1999a} relation are those with the highest average IR flux among their analogs, in agreement with \cite{Howell2010} ($z\sim0$ and \cite{Penner2012} ($z\sim2$). 
However, many of these studies used UV- or IR-selected samples, whereas our work uses the largest sample yet of mass-selected galaxies at $z\sim2$.

More recently, \cite{Talia2015} analyzed the $A_{1600} - \beta$ relation using high-redshift UV spectra of 62 IR-detected galaxies at $1<z<3$, obtaining a much flatter fit than previous work, although still broadly consistent with predictions based on the Calzetti attenuation law.
Our work, utilizing photometry only, includes a much larger sample, suggesting that previous works making use of local IRX$-\beta$ relations incorrectly estimate the extinction of the UV continuum for high redshift dust-obscured samples.

This offset implies that the dust attenuation at redshifts $z\sim1-3$ is different from that in local galaxies; specifically, the steeper slope of our $A_{1600}-\beta$ relation means that dusty star-forming galaxies in our sample's redshift range have more UV attenuation due to dust than would be assumed from local calibrations.
This amounts to a 0.5 magnitude underestimate of 1600 $\textrm{\AA}$ attenuation for galaxies with $\beta=0$, increasing towards $\beta\sim1$, and becoming consistent with the M99 relation for $\beta\sim-1$.

\subsection{Dust Attenuation from Composite SED Colors} \label{Res3}

Both the IRX and \textit{UVJ} colors separate the red and blue star-forming populations effectively; therefore we can also analyze the relation between IRX and these colors.
We fit our $A_{1600}$ values derived from the IRX$-\beta$ fits to the rest-frame colors (Figure \ref{fig: L_A1600}).
The resulting linear relations are
\begin{eqnarray}
A_{1600} &=& (3.64 \pm 0.23) (U-V)_{REST} - (0.43 \pm 0.24)\\
A_{1600} &=& (2.86 \pm 0.30) (V-J)_{REST} + (0.58 \pm 0.30),
\end{eqnarray}
which can be used in conjunction with the IRX$-A_{1600}$ relation to obtain
\begin{eqnarray}
\textrm{IRX} &=& 1.68\times [10^{0.4(3.64 (U-V)_{REST} - 0.43)}-1]\\
\textrm{IRX} &=& 1.68\times [10^{0.4(2.86 (V-J)_{REST} + 0.58)}-1].
\end{eqnarray}
Our derived $A_{1600}$ for star-forming galaxies correlates well with the $(U-V)_{REST}$ color, as can be seen in the upper central panel of Figure \ref{fig: L_A1600}, allowing it to be used as a proxy for UV dust attenuation at these redshifts for all but the reddest star-forming galaxies. 
Because rest-frame colors are fairly easily determined, these relations provide a useful way to estimate dust corrections for star-forming galaxies without requiring a spectrum.


	\begin{table}[h]
	\caption{Fit parameters of the IRX$-\beta$ relation.$^a$}
	\begin{tabular}{c || r r r r l}
	Paper & q & $\delta$q & r & $\delta$r & Sample \\ \hline
 	This Work		& 5.05	& 0.16	& 2.39	& 0.14	&  $1<z<3$ composite SEDs \\
	Meurer 1999	& 4.43	& 0.08	& 1.99 	& 0.04	& local starbursts\\
	Penner 2012	& 5.94	& --	  	& 2.34	& --		& $z\sim2$ DOGs, 24 $\mu m$-selected  (61)\\
	Casey 2014	& 3.36	& 0.10	& 2.04	& 0.08	& $z<0.085$, IRX$<$60\\
	Talia 2015		& 3.33	& 0.24	& 1.10	& 0.23	& $1<z<3$ SFG spectra (62)
	\label{table:comp}
	\footnotetext{$IRX = 1.68(10^{0.4(q+r\beta)}-1)$}
	\end{tabular}
	\end{table}


We use the above relations to derive the direction of the unit vector of IRX and $A_{1600}$ on the \textit{UVJ} diagram.
While the IRX trend more or less parallels the distribution of the star-forming composite SEDs, it is apparent that the $A_{1600}$ vector on the \textit{UVJ} diagram does not parallel the color evolution of star-forming galaxies as $A_V$ does \citep{Wuyts2007, Williams2009}.

\section{Conclusions} \label{Conc}

In this work we empirically generate a set of 38 composite SEDs from the photometric data points of $\sim$4000 $K_s$-band selected galaxies in the redshift range $1<z<3$ from the mass complete ZFOURGE survey.
The use of composite SEDs allows us to densely probe the SEDs of large samples of galaxies without spectroscopy.
We analyze these galaxies using the \textit{UVJ} diagnostic plot, and verify that this simple color$-$color relation test does an excellent job of separating quiescent galaxies from star-forming galaxies.

In addition, we explore the IRX$-\beta$ relation, which parameterizes the dust content of galaxies, for our star-forming composite SEDs.
While there is a range of results regarding IRX$-\beta$ in previous studies, this work utilizes a large, mass-selected sample to derive the relation at intermediate redshifts.
We find that these composite SEDs lie above the relations derived for local samples of galaxies, indicating differences in dust properties.
Specifically, dusty star-forming galaxies have more dust attenuation in the UV, and are therefore intrinsically bluer, than would be derived based on calibrations from samples of non-dust obscured galaxies extending the results of previous work in the local universe to $z\sim2$.

We also find trends with low scatter between dust attenuation and rest-frame \textit{UVJ} colors for star-forming galaxies, which can be used to parameterize dust of star-forming galaxies without spectra.
Future work will explore other properties of our composite SEDs, such as H$\alpha$ equivalent widths and morphological characteristics.

\section*{Acknowledgments}
We wish to thank the Mitchell family, particularly the late George P. Mitchell, for their continuing support of astronomy.
We also thank the Carnegie Observatories and the Las Campanas Observatory for their assistance in making the ZFOURGE survey possible.
Additionally, thanks to the CANDELS-\textit{Herschel} team for allowing us to use the unpublished PACS 100 $\mu m$ and 160 $\mu m$ data for the COSMOS and UDS fields, which helped constrain the IR fluxes of our composite SEDs.
Finally, we are grateful to Caitlin Casey and Brett Salmon for several great conversations regarding the IRX$-\beta$ relation and dust attenuation, and Kyle Penner for providing improvements to the manuscript.
K. Tran acknowledges the support of the National Science Foundation under Grant $\#$1410728.
GGK is supported by an Australian Research Council Future Fellowship FT140100933.


\end{document}